\documentclass[twocolumn,showpacs,preprintnumbers,amsmath,amssymb,pra]{revtex4}
                                                                                
\usepackage{graphicx}
\usepackage{dcolumn}
\usepackage{color}
\usepackage{bm}
\begin{document}

\title {\bf Observability of the arrival time distribution using 
spin-rotator as a quantum clock}

\author{Alok Kumar Pan\footnote{apan@bosemain.boseinst.ac.in}$^1$,
Md. Manirul Ali\footnote{mani@bose.res.in}$^2$, and
Dipankar Home\footnote{dhome@bosemain.boseinst.ac.in}$^1$}

\address{$^1$Department of Physics, Bose Institute, Calcutta
700009, India}

\address{$^2$S. N. Bose National Centre for Basic Sciences,
Block JD,
Sector III, Salt Lake, Calcutta 700098, India}

\date{\today}
                                                                                
\begin{abstract}
                                                                                 
An experimentally realizable scheme is formulated which can test \emph{any}
postulated quantum mechanical approach for calculating the arrival
time distribution. This is specifically illustrated by using the modulus
of the probability current density for calculating the arrival time
distribution of spin-1/2 neutral particles at the exit point of a
spin-rotator(SR) which contains a constant magnetic field. Such a
calculated time distribution is then used for evaluating the distribution
of spin orientations along different directions for these particles
emerging from the SR. Based on this, the result of spin measurement
along any arbitrary direction for such an ensemble is predicted. 
\end{abstract}        

\pacs{03.65.Bz}
\maketitle                                                                                 

Introduction.---Of late, the question of calculating
the arrival or transit time distribution in quantum mechanics has
been a topic of much interest. For comprehensive reviews see, for
example, Muga and Leavens \cite{1}, and Muga et al. \cite{2}. 
A number of schemes \cite{3,4,5}
have been suggested in the literature for calculating
the arrival time distribution such as those based on axiomatic approaches,
trajectory models of quantum mechanics, attempts to define and calculate
the arrival time distribution using the consistent histories approach,
and attempts of constructing the time of arrival operator, etc.
Thus there is an inherent \emph{nonuniqueness} within the
formalism of quantum mechanics for calculating time distributions
such as the arrival time distribution.\\ 
                                                                                        
Against the backdrop of such studies, it remains an open question
as to what extent these different quantum mechanical approaches for
calculating the time distributions can be empirically discriminated.
An effort along this direction was made by Damborenea et al. \cite{6}
who considered the measurement of arrival time by the emission of
a first photon from a two-level system moving into a laser-illuminated
region. They had evaluated the probability for this emission of the
first photon by using the quantum jump approach. The suitable approximations
under which such calculated results could be related to Kijowski's
axiomatic arrival time distribution and the arrival time distribution
defined in terms of the probability current density (\emph{not} its
modulus) were also discussed. Subsequently, further work was done
along this direction by Hegerfeldt et al. \cite{7} who made more precise
the connection of this approach with Kijowski's distribution. \\

In this paper we address this question from a new perspective so that
one can start from an axiomatically defined time distribution and
then \emph{directly} relate it to the actually testable results. In
order to illustrate this approach, here in particular, we use a time
distribution \emph{postulated} in terms of the (normalised) \emph{modulus}
of the \emph{probability current density}. Based on this, we derive
a distribution of spin orientations along different directions for
the spin-1/2 neutral particles emerging from a spin-rotator (SR) which
contains a constant magnetic field. Such a calculated distribution
function can then be tested by suitably using a Stern-Gerlach(SG)
device, as explained later. Thus the scheme formulated in this paper
can \emph{also} be viewed as a verification of the \emph{observability}
of the \emph{quantum probability current density}.\\
                                                                                                      
Unlike the position probability density, the status of probability
current density in quantum mechanics as an observable quantity has
remained a problematic issue; see, for example, Kan and Griffin \cite{8}
who pointed out that for a many-particle system, any linear operator
representation for velocity is inconsistent with a linear operator
representation for the probability current density, such as the one
constructed by Landau \cite{9} for the quantum theory of superfluid
helium. Nevertheless, in the context of single-particle dynamics,
the probability current density has been used in the quantum mechanical
predictions of time distributions such as the arrival time \cite{4,5,10},
tunneling and reflection times \cite{11}.\\
                                                                                        
Next, let us consider the analysis of the experimental techniques
for measuring the {}``arrival time'' or {}``time of flight''.
We note that such analysis is usually done semi-classically or classically
\cite{12}. Therefore it is curious that the question of a consistent
quantum mechanical treatment of the measurement of time has remained
murky ever since Pauli's argument \cite{13} that {}``there cannot
be a self adjoint time operator conjugate to any Hamiltonian bounded
from below''. Subsequently, a number of authors \cite{14} have pointed
out various conceptual and mathematical problematic aspects of this
question. On the other hand, several specific toy models \cite{15}
have also been proposed to investigate the feasibility of \emph{how}
actually the measurement of a time distribution can be performed in
a way consistent with the basic principles of quantum mechanics. \\
                                                                                        
Now, since such debates arise essentially if one considers \emph{how}
to \emph{directly measure} time in quantum mechanics, here we bypass
this vexed issue by adopting the following strategy. We consider the
SR as a \emph{{}``quantum clock''} where the basic quantity which
determines the actually observable results is the probability density
function \( \Pi \left( \phi \right)  \) which corresponds to the
probability distribution of spin orientations along different directions
for the particles emerging from the SR, \( \phi  \) being the angle
by which the spin orientation of a spin-1/2 neutral particle (say,
a neutron) is rotated from its initial spin polarised direction. Note
that this angle \( \phi  \) is determined by the transit time \( (t) \)
within the SR. Hence the probability density function \( \Pi \left( \phi \right)  \)
stems from \( \Pi \left( t\right)  \) which represents the distribution
of times over which the particles interact with the constant magnetic
field while passing through the SR. It is the evaluation of this quantity
\( \Pi \left( t\right)  \) which critically depends on what quantum
mechanical approach one adopts for calculating such a time distribution.\\
                                                                                        
The plan of this paper is as follows. We will first elaborate on the
relevant setup, with a discussion of how the estimation of the quantity
\( \Pi \left( \phi \right)  \) can actually be tested by using a
SG device. Subsequently, we will outline the key ingredients of the
specific scheme adopted in this paper for calculating \( \Pi \left( \phi \right)  \)
in terms of \( \Pi \left( t\right)  \) using the modulus of the probability
current density. This will be followed by illustrative numerical estimates.\\

The setup.---Traditionally, a SR has been mainly used for the
neutron interferometric studies \cite{16}. Application of the Larmor
precession of spin in a magnetic field has earlier been discussed,
for example, in the context of the scattering of a plane wave from
a potential barrier \cite{17}. On the other hand, the scheme proposed
in this paper explores an application of Larmor precession such that
one can empirically test \emph{any} given quantum mechanical formulation
for calculating the arrival time distribution.\\

We consider an ensemble of spin \( 1/2 \) neutral particles, say,
neutrons having magnetic moment \( \mu  \). The spatial part of the
total wave function is represented by a localised narrow Gaussian
wave packet \( \psi \left( x,t=0\right)  \) (for simplicity, it is
considered to be one dimensional) which is peaked at \( x=0 \) at
\( t=0 \) and moves with the group velocity \( u \). Thus the initial
total wave function is given by \( \Psi =\psi \left( x,t=0\right) 
\otimes \chi \left( t=0\right)  \)
where \( \chi \left( t=0\right)  \) is the initial spin state which
is taken to be \emph{same} for all members of the ensemble.\\
                                                                                        
The SR used in our setup (Fig. 1) has within it a constant magnetic
field \( {\bf B}=B\widehat{\bf z} \) directed along the \( +\widehat{\bf z} \)
-- axis, confined between \( x=0 \) and \( x=d \). Within the SR,
the spatial part of the total wave function is assumed to propagate
freely, while its spin part interacts with the constant magnetic field.
This assumption is justified in our setup because, for our choices
of parameters, the magnitude of the Zeeman potential energy of the
interaction of spin of the neutron with the constant magnetic field
(\( \simeq 0.01\, neV \)) is exceedingly
\emph{small} compared to the kinetic energy of the neutrons (\( \simeq 0.01\, eV \)).
Hence to a very high degree of accuracy we can consider the evolution
of the spatial wave function within the SR to be \emph{free.} Therefore
the spatial and the spin parts of the total wave function can be considered
to evolve independent of each other in a tensor product Hilbert space
\( H=H1\otimes H2 \) where \( H1 \) and \( H2 \) are the disjoint
Hilbert spaces corresponding to the spatial and the spin parts of
the total wave function respectively \cite{16}. \\

Here in our analysis we assume that the spin part of the total wave
function \emph{begins} to interact with the constant magnetic field
in the SR at the instant \( \left( t=0\right)  \) \emph{when} the
peak of the incoming wave packet is at the entry point \( \left( x=0\right)  \)
of the SR. Thus the calculational procedure adopted here is esssentially
valid for a \emph{sufficiently narrow} wave packet. Then it can be
assumed that the \emph{entire} ensemble of particles corresponding
to the initial wave packet \emph{start} interacting with the magnetic
field at \( t=0 \). This assumption is crucial in this scheme in
order to enable the arrival / transit time distribution \( \Pi \left( t\right)  
\) to be mapped onto \( \Pi \left( \phi \right)  \). 

It is important to mention here that in the standard approach, the question 
as regards the initial {\it instant} at which a {\it propagating wave packet} 
of any arbitrary width starts interacting with a localized potential ({\it e.g.} 
the localized magnetic field in our setup) is intrinsically problematic since 
there is no unique criterion for fixing this instant. Implications of this 
nonuniqueness have hitherto remained unaddressed in the literature. Following 
the usual procedure (expected to be valid for narrow wave packets) we have assumed 
in our paper that the spin part of the wave function {\it begins} to evolve under 
a given localized potential essentially from the instant {\it when} the peak of 
the wave packet reaches the entry point of that potential. This criterion is of 
course not the only criterion one can use to define this initial instant and one 
can certainly use any other criterion to fix this {\it initial instant}.

Basically there are {\it two} different kinds of {\it non-uniqueness} in the quantum 
mechanical treatment of our problem; one is that there is {\it no} unique criterion 
available within the standard framework of quantum mechanics for fixing the initial 
instant {\it when} the spin part of the wave function {\it starts} to interact with 
the localized potential; the other {\it non-uniqueness} is inherent within the 
formalism of quantum mechanics is regarding the {\it time-duration} over which the 
wave packet interacts with the localized potential which in our paper is fixed by 
the {\it arrival/transit time distribution}. These are the crucial conditions relevant 
to our work. 

The important point is that using our scheme one can {\it test} experimentally {\it any}
postulated quantum mechanical approach for calculating the arrival time distribution 
with different criteria for fixing the initial time of interaction. Note that 
a number of approaches have been suggested in the literature for calculating the arrival 
time distribution. We have adopted in our paper one particular approach, viz., the 
{\it current density} approach to calculate the arrival time distribution using a particular  
method of calculation ( {\it e.g.} fixing the initial time of interaction in terms of the 
peak), as an example, to illustrate our {\it general scheme}.

\begin{figure}[t]
{\rotatebox{0}{\resizebox{8.0cm}{3.5cm}{\includegraphics{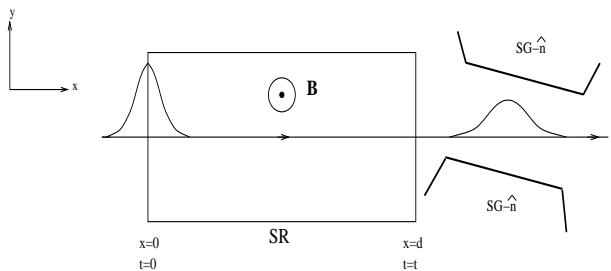}}}}
\caption{\label{fig.1} {\footnotesize Spin-1/2 particles, say, neutrons
with initial spin orientations polarised along the \( +\widehat{\bf x} \)
- axis and associated with a localized Gaussian wave packet (peaked
at \( x=0 \), \( t=0 \)) pass through a spin-rotator (SR) containing
a constant magnetic field \( {\bf B} \) directed along the \( +\widehat{\bf z} \)
- axis. The particles emerging from the SR have a distribution of
their spins oriented along different directions. Calculation of this
distribution function is experimentally tested by measuring the spin
observable along a direction \( \widehat{n}\left( \theta \right) \)
in the xy-plane making an angle \( \theta  \) with the initial spin polarised along
\( +\widehat{\bf x} \)- axis. This is done by suitably orienting the direction
\( \widehat{n}\left( \theta \right)  \) of the inhomogeneous magnetic
field in the Stern-Gerlach \( \left( SG-\widehat{n}\right)  \) device.}}
\end{figure}

Next, we recall that when a spin-polarised particle (say, a neutron)
passes through the constant magnetic field within a SR, its spin orientation
is rotated by an angle \( \phi  \) with respect to the initial spin
polarised direction along \( +\widehat{\bf x} \) axis. This angle is
fixed by the time \( (t) \) spent by the particle within the SR,
given by the well known quantum mechanical relation \( \phi =2\omega t \)
where \( \omega =\mu B/\hbar  \) \cite{18}. \\

Now, let us consider an ensemble of particles, say, neutrons passing through 
the SR where initially \emph{all} members of this ensemble have their spins 
polarised along, say, the \( +\widehat{\bf x} \) axis. Given the same initial
spin state they evolve over different times (characterised by \( \Pi \left( t\right) \)
, the \emph{distribution} \emph{of transit times} within the SR) under the interaction 
with the constant magnetic field within the SR for many repetitions of the experiment.

Here we would like to stress that in our setup, the very fact that
a distribution of spins emerges from the SR implies the existence
of a distribution of transit times within the SR, since the
spin rotation is proportional to the transit time. Hence in order
that a distribution of spins emerges from the SR, the spin
part of the wave function going through the SR must
necessarily interact with the SR magnetic field over \emph{different
times} in each run of the experiment. It then follows that the unitary 
evolution operators $U=\exp(-iHt/\hbar)$ are \emph{different} for each 
repetition of the experiment, although the spin interaction Hamiltonian \( H \) 
is the \emph{same} for \emph{all} of them. Thus the emergent spin states 
get polarised along different
directions and consequently the final ensemble of particles emerging
from the SR is in a \emph{mixed state} of spin states polarised along
various directions with different respective probabilities. \\

Hence we can write the final density matrix of the total ensemble at
{\it any} time which is large enough so that by which all the particles
of the ensemble (total wave packet) have passed through the spin rotator
(SR) to be given by

\begin{equation}
\label{1}
W_f=\sum _{t}\Pi \left( t\right) \left| \chi \left( t\right) \right\rangle \left\langle \chi \left( t\right) \right|
\end{equation}

where $|\chi(t)\rangle$'s  occurring in the right
hand side of Eq.(1) are the time evolved pure states which have evolved under
the given potential within SR over different times (denoted by the symbol ``t'').
The final density matrix by combining these time evolved spin states is 
written at a sufficiently large time by which all the members of the ensemble
have passed through the SR. Now, using the relation \( t=\phi /2\omega  \), 
one can rewrite the density matrix given by Eq.(1) in the following form 
\begin{equation}
\label{2}
W_f=\sum _{\phi }\Pi \left( \phi \right) \left| \chi \left( \phi \right) \right\rangle \left\langle \chi \left( \phi \right) \right|
\end{equation}

where the summation in Eq. (2) is over the different values of \( \phi  \)
corresponding to different values of transit time \( \left( t\right)  \)
within the SR, and \( \left| \chi \left( \phi \right) \right\rangle  \)
is the normalised spin state which represents the spin polarization
along any direction making an angle \( \phi  \) with the \( +\widehat{x} \)
axis. Here \( \Pi \left( \phi \right)  \) is the normalised probability
density of spin orientations which is obtained from \( \Pi \left( t\right)  \)
through the relation \( t=\phi /2\omega  \). The quantity \( \Pi \left( \phi 
\right) d\phi  \) represents the probability of spins emerging from the SR 
having their orientations within the angles \( \phi  \) and \( \phi +d\phi  \).
Note that using Eq.(2), it follows that
\begin{eqnarray}
\nonumber
Tr \{ W_f \}&=&\sum_i \langle u_i|W_f|u_i \rangle=
\sum_{\phi}\Pi(\phi)\sum_i |\langle u_i|\chi(\phi)\rangle|^2\\
&=& \sum_{\phi} \Pi(\phi)={\bf 1}
\end{eqnarray}
where we've introduced the states \( \left\{ \left| u_{i}\right\rangle 
\right\}  \) as a complete set of orthonormal basis for any spin state 
\( \left| \chi \left( \phi \right) \right\rangle  \), and hence \( \sum _{i}
\left| \left\langle u_{i}\mid \chi \left( \phi \right) \right\rangle \right| 
^{2}=1 \). Note that the above result is valid regardless of whether the spin
states \( \left| \chi \left( \phi \right) \right\rangle  \) for the
different values of \( \phi  \) are orthogonal or not.\\
                                                                                        
Testability of \( \Pi \left( \phi \right)  \) using the Stern-Gerlach
device.---Now, for testing the scheme we have outlined for calculating
the probability density function \( \Pi \left( \phi \right)  \),
let us consider the measurement of a spin variable, say \( \widehat{\sigma }_{\theta } \),
by a SG device (Fig.1) in which the inhomogeneous magnetic field
is oriented along a direction \( \widehat{n}\left( \theta \right)  \) in
the xy-plane making an angle \( \theta  \) with the initial spin-polarised direction
(\( +\widehat{\bf x} \) axis) of the particles. Then for the spins of
the particles emerging from the SR polarised along different directions
(with the probabilities \( \Pi \left( \phi \right)  \)) making angles
\( \phi  \) with the \( +\widehat{\bf x} \) axis, the probabilities
of finding the spin component along \( +\theta  \) direction and
that along its opposite direction are respectively given by\\
\begin{equation}
\label{4}
P_{+}\left( \theta \right) =\int_{0}^{2\pi }\Pi \left( \phi \right)
{Cos}^2 \frac{\left( \theta -\phi \right)}{2} d\phi
\end{equation}
\begin{equation}
\label{5}
P_{-}\left( \theta \right) =\int_{0}^{2\pi }\Pi \left( \phi \right)
{Sin}^2 \frac{\left( \theta -\phi \right)}{2} d\phi
\end{equation}
where \( P_{+}\left( \theta \right) +P_{-}\left( \theta \right) =1 \),
and here we are essentially restricting to the situations in which
the relevant parameters \( d \), \( u \) and \( B \) are such that
the spin rotation angles \( \phi  \) for all the particles emerging
from the SR are restricted between \( \phi =0 \) and \( \phi =2\pi  \).
To explain in more detail how Eqs.(4) and (5) are derived, let us first
consider particles passing through the spin rotator (SR) with all their
spins oriented along a definite direction, say, x-axis. The initial 
x-polarised spin state can be written in terms of the z-bases 
$|\uparrow {\rangle}_z$ and $|\downarrow {\rangle}_z$ as 
$\chi (0)=1/{\sqrt{2}}~ \left(|\uparrow {\rangle}_z +|\downarrow {\rangle}_z \right)$.
Then in such a case, the spin polarised state rotates only in the xy-plane.
If $\phi$ is the rotation angle with respect to the initial spin orientation
along $\widehat{\bf x}$-axis, such a rotated spin state in the xy-plane can be typically 
written as $\chi (\phi)=1/{\sqrt{2}}~\left(|\uparrow {\rangle}_z + e^{i \phi}
|\downarrow {\rangle}_z \right)$. Now, for the purpose of measurement after the 
spins emerge from SR, if one applies SG-magnetic field along a direction
\( \widehat{n}\left( \theta \right)  \) in the xy-plane making an angle 
$\theta$ with the $\widehat{\bf x}$-axis, then the bases states for the spin operator 
${\hat \sigma_{\theta}}$ are respectively $|\uparrow \rangle_{\theta}
=1/{\sqrt{2}}~\left(|\uparrow 
{\rangle}_z + e^{i \theta}|\downarrow {\rangle}_z \right)$ and $|\downarrow 
\rangle_{\theta}=1/{\sqrt{2}}~\left(|\uparrow {\rangle}_z - e^{i \theta} 
|\downarrow {\rangle}_z \right)$. Then for this spin measurement the probabilities 
of getting 
$|\uparrow \rangle_{\theta}$ and $|\downarrow \rangle_{\theta}$ are
$p_{+}(\theta)={|}_{\theta}{\langle \uparrow}~|~\chi(\phi)\rangle|^2=Cos^2 (\theta-\phi)/2$
and $p_{-}(\theta)=~{|}_{\theta}{\langle \downarrow}~|~\chi(\phi)\rangle|^2=Sin^2 (\theta-\phi)/2$
respectively. Now, since in our setup, instead of a definite spin polarised state,
we have considered a distribution of spin orientations along different directions
for the particles emerging from the SR characterised by the distribution 
function $\Pi(\phi)$, consequently we have the expressions for the probabilities
$P_{+}(\theta)$ and $P_{-}(\theta)$ as given by Eq.(4) and Eq.(5) respectively.
Note that the departures obtained from the prediction given by semiclassical
approach (which assumes that, neglecting the effect due to the spreading
of the wave packet, the spins of \emph{all} members of the ensemble
rotate by an amount which is determined by the time spend by the peak
of the wave packet in traversing the region within the SR) can be observed
by sensitive measurements of \( P_{+}\left( \theta \right)  \) and
\( P_{-}\left( \theta \right)  \). \\

It is these probabilities \( P_{+}\left( \theta \right)  \) and 
\( P_{-}\left( \theta \right)  \)
which constitute the basic \emph{observable quantities} in this scheme
which are determined by the distribution of spins \( \Pi \left( \phi \right)  \)
of the particles emerging from the SR. The estimations of these probabilities
crucially depend on \emph{how} one calculates the quantity \( \Pi \left( \phi \right)  \)
whose evaluation, in turn, is contingent on the procedure adopted
for calculating the relevant time distribution \( \Pi \left( t\right)  \).
As mentioned earlier, the specification of such a time distribution
is \emph{not} unique in quantum mechanics. For the setup indicated
in Fig.1, \( \Pi \left( t\right)  \) represents the \emph{arrival
time distribution} at the exit point \( \left( x=d\right)  \) of
the SR, which is \emph{also} the \emph{distribution of transit times}
\( \left( t\right)  \) within the SR. In the specific scheme we are
using, \( \Pi \left( t\right)  \) is taken to be represented by the
modulus of the probability current density \( \left| {\bf J}\left( {\bf X},t\right) \right| \) 
(suitably normalised) evaluated at the spatial point {\bf X}(x=d,y=0,z=0); i.e., we take 
$\Pi(t)=|{\bf J}({\bf X},t)|/\int_0^\infty |{\bf J}({\bf X},t)|dt$.
Leavens \cite{4,5} has justified the above interpretation of the modulus of probability
current density $|{\bf J}({\bf X},t)|$ as an (unnormalized) arrival time distribution
using the Bohm's causal model of quantum mechanics. 
Hence $\Pi(\phi)=|{\bf J}({\bf X},\phi)|/\int_0^{2\pi} |{\bf J}({\bf X},\phi)|d\phi$.
Thus in this scheme, the calculation of $\Pi(\phi)$ ultimately hinges on evaluating 
${\bf J}({\bf X},\phi)$ from ${\bf J}({\bf X},t)$. \\

The evaluation of $\Pi(\phi)$.---Since we have to first calculate ${\bf J}({\bf X},t)$,
we begin by recalling that the standard expression for the non-relativistic
quantum probability current density is given by\\
\begin{equation}
\label{6}
{\bf J}({\bf x},t) =Re\left[ \psi ^{*}\left({\bf x},t\right) \left( -\frac{i\hbar }{m}\right) 
{\bf \nabla }\psi \left({\bf x},t\right) \right]
\end{equation}
which satisfies the quantum mechanical equation of continuity given by
\begin{equation}
\label{7}
\frac{\partial \rho }{\partial t}+{\bf \nabla }.{\bf J}=0
\end{equation}
where the position probability density \( \rho \left({\bf x},t\right) 
=\psi ^{*}\left({\bf x},t \right) \psi \left({\bf x},t\right)  \).\\ 

Then comes a key point. If one adds any \emph{divergence-free term} to the above 
expression for \( {\bf J}({\bf x},t) \), then
the new expression also satisfies the same equation of continuity.
Hence there is a \emph{nonuniqueness} inherent in the nonrelativistic
expression for the probability current density. Curiously, this point
has not been noted even in the premier textbooks like that by Landau
and Lifshitz \cite{19}, Merzbacher \cite{20}. However, relatively recently
this problem of nonuniqueness has been highlighted \cite{21} and it
has been pointed out that the probability current density derived
from the Dirac equation for any spin-1/2 particle is \emph{unique}
and even in the \emph{non-relativistic limit} it contains a \emph{spin-
dependent term} which is present in addition to the expression for
\( {\bf J}({\bf x},t) \) given by Eq.(6). Interestingly,
one can further argue that this property of the \emph{uniqueness}
of probability current density is \emph{not} specific to the Dirac
equation, but is a consequence of \emph{any} relativistic quantum
mechanical equation. The argument is as follows.\\
                                                                                        
The probability current density obtained from \emph{any} consistent
relativistic quantum mechanical equation needs to satisfy a covariant
form of the continuity equation of \( j^{\mu } \) where the zeroth
component of \( j^{\mu } \)\( (j^{0}) \) is associated with the
position probability density. If one replaces \( j^{\mu } \) by \( \overline{j}^{\mu } \)
which is also conserved, i.e., \( \partial _{\mu }\overline{j}^{\mu }=0 \)
where \( \overline{j}^{\mu }=j^{\mu }+a^{\mu } \) (\( a^{\mu } \)
is an arbitrary 4-vector), then the zeroth component of \( \overline{j}^{\mu } \)
\( (\overline{j}^{0}) \) will have to be the \emph{same} as the position
probability density given by \( j^{0} \). Hence it follows that \( a^{0}=0 \).\\

Next, we consider this current as seen from another Lorentz frame.
This is given by \( \overline{j}^{\mu \prime }=j^{\mu }+a^{\mu \prime } \).
Hence in this frame \( \overline{j}^{0\prime }=j^{0}+a^{0\prime } \),
and again if the position probability density has to remain \emph{unchanged},
then one must have \( a^{0\prime }=0 \). But we know that the \emph{only}
4-vector whose fourth component vanishes in \emph{all} frames is the
\emph{null vector}. Thus \( a^{\mu }=0 \). It therefore follows that
for any consistent relativistic quantum mechanical equation satisfying
the covariant form of the continuity equation, the relativistic current
is \emph{uniquely fixed}. This uniqueness is \emph{also} preserved
in the non-relativistic limit of the relevant relativistic equation.
\\

Now, in order to make this paper self-contained, we briefly recapitulate
the key steps involved in deriving the expression for the probability
current density in the non-relativistic limit from the Dirac equation
in \( 3+1 \) dimension for a \emph{free} spin-1/2 particle of rest
mass \( m_{0} \) given by
\begin{eqnarray}
i\hbar\frac{\partial \Psi}{\partial t} = \left[\frac{\hbar c}{i}\hskip 0.2cm {\alpha}_i \frac{\partial}{\partial x_i}+\beta m_0 c^2 \right]\Psi
\end{eqnarray}
where
\[\alpha_i = \left (\begin{array}{cc}
0 & \sigma_i\\
\sigma_i & 0 \\
\end{array}\right),\beta = \left (\begin{array}{cc}
I & 0\\
0 & -I \\
\end{array}\right),\Psi = \left (\begin{array}{c}
\Psi_1\\
\Psi_2\\
\end{array}\right)\] 
and $\Psi_1$, $\Psi_2$ are individually two component spinors.
Subsequently, Eq. (8) leads to two coupled equations\\
\begin{equation}
\label{9}
\frac{\partial \Psi _{1}}{\partial t}=-c\sigma _{i}\frac{\partial \Psi _{2}}{\partial x_{i}}-\frac{i}{\hbar }m_{0}c^{2}\Psi _{1}
\end{equation}
\begin{equation}
\label{10}
\frac{\partial \Psi _{2}}{\partial t}=-c\sigma _{i}\frac{\partial \Psi _{1}}{\partial x_{i}}+\frac{i}{\hbar }m_{0}c^{2}\Psi _{2}
\end{equation}
Then taking the positive energy solution \( \Psi _{2}\propto \exp \left( -iEt/\hbar \right)  \),
substituting it in Eq.(10) and putting \( E\cong m_{0}c^{2} \) in
the \emph{non-relativistic regime}, we get\\
\begin{equation}
\label{11}
\Psi _{2}=-\frac{i\hbar }{2m_{0}c}\sigma _{i}\frac{\partial \Psi _{1}}{\partial x_{i}}
\end{equation}
Multiplying Eq. (9) by \( \Psi ^{\dagger }_{1} \) from the left and
multiplying again the hermitian conjugate of Eq. (9) by \( \Psi _{1} \)
from the right, we add the two equations. Substituting the value of
\( \Psi _{2} \) from Eq.(11) in this resulting equation, one can
then obtain the following equation given by
\begin{eqnarray} 
&&\frac{\partial}{\partial t}({\Psi_1}^\dagger {\Psi_1})+\\
\nonumber
&&\frac{\partial}{\partial x_i}\left[ -\frac{i \hbar}{2 m_0}
\left\{ {\Psi_1}^\dagger \sigma_i\left(\sigma_i
\frac{\partial \Psi_1}{\partial x_i} \right)
- \left( \frac{\partial {\Psi_1}^\dagger}{\partial x_i} 
\sigma_i \right) \sigma_i \Psi_1    \right\} \right]\\
\nonumber
&&=0
\end{eqnarray}
Now, comparing Eq.(12) with Eq.(7), it is seen that the Dirac current
in \( 3+1 \) dimension for a \emph{free} spin-1/2 neutral particle
in the \emph{non-relativistic limit} is of the form given by 
\begin{eqnarray}
&&{\bf J}\left({\bf x},t\right)\\
\nonumber
&=&\left[ -\frac{i\hbar }{2m_{0}}\left\{ \Psi ^{\dagger }_{1}\sigma _{i}\left( \sigma _{i}\frac{\partial \Psi _{1}}{\partial x_{i}}\right) -\left( \frac{\partial \Psi ^{\dagger }_{1}}{\partial x_{i}}\sigma _{i}\right) \sigma _{i}\Psi _{1}\right\} \right]
\end{eqnarray}
where \( \Psi _{1} \) is a two component spinor which can be written
as \( \Psi _{1}=\psi \left({\bf x},t\right) \chi \left( t\right)  \).
Simplifying Eq.(13) by using the Gordon decomposition \cite{21,22},
one finally obtains                                                             \begin{eqnarray}
&&{\bf J}\left({\bf x},t\right)\\
\nonumber
&&=Re\left[ \psi ^{*}\left({\bf x},t\right) \left( -\frac{i\hbar }
{m_{0}}\right) {\bf \nabla }\psi \left({\bf x},t\right) \right] +
\frac{1}{m_{0}}\left[ {\bf \nabla }\rho \times {\bf s}\left( t\right) \right]\\
\nonumber
&&={\bf J}_{Sch}\left({\bf x},t\right) +{\bf J}_{Spin}\left({\bf x},t\right) 
\end{eqnarray}
where $\rho=\left|\psi\left({\bf x},t\right)\right|^{2}$,
${\bf s}(t)=\frac{\hbar}{2}\chi^{\dagger}(t){\bm \sigma}\chi(t)$, ${\bm \sigma}
=\sigma_{x}\widehat{\bf x}+\sigma_{y}\widehat{\bf y}+\sigma_{z}\widehat{\bf z}$
and $\chi(t)^{\dagger}\chi(t)=1$. In Eq.(14), the first term represents the 
usual Schroedinger current $\left({\bf J}_{Sch}({\bf x},t)\right)$ and the second
term gives the contribution of the spin dependent current 
$\left({\bf J}_{Spin}({\bf x},t)\right)$. The above decomposition is possible
because there is no spatial dependence on the spin state $\chi(t)$ which is only
time dependent here in our case.  

Next, in the context of our setup, in order to evaluate the quantity
\({\bf J}\left( x={\bf X}, \phi \right)  \), we
first consider the spatial part of the total wave function. As mentioned
earlier, it is taken to be a \emph{one dimensional} Gaussian wave
packet which is peaked at the entry point \( \left( x=0\right)  \)
of the SR at \( t=0 \) (Fig. 1). The initial spatial wave function
is then given by 
\begin{equation}
\label{15}
\psi \left( x,t=0\right) =\frac{1}{\left( 2\pi \sigma ^{2}_{0}\right) ^{1/4}}\exp \left[ \frac{-x^{2}}{4\sigma ^{2}_{0}}+ikx\right]
\end{equation}
where \( \sigma _{0} \) is the initial width of the associated wave
packet. The wave number \( k=m_{0}u/\hbar  \) where \( u \) is the
group velocity of the wave packet moving along the \( +\widehat{\bf x} \)
-- axis. Since the spatial part of the total wave function propagates
freely, being unaffected by the constant magnetic field confined within
the SR, the Schroedinger time evolved spatial wave function calculated
from the initial wave function \( \psi \left( x,t=0\right)  \) given
by Eq. (15) is of the form
\begin{equation}
\label{16}
\psi \left( x,t\right) =\frac{1}{\left( 2\pi A^{2}_{t}\right) ^{1/4}}\exp \left[ -\frac{( x-ut)^2}{4 A_t \sigma_0}+ik(x-\frac{1}{2}ut) \right]
\end{equation}
and hence the time evolved position probability density is given
by
\begin{equation}
\label{17}
\rho \left( x,t\right) =\frac{1}{\left( 2\pi \sigma ^{2}_{t}\right) ^{1/2}}\exp \left[ -\frac{\left( x-ut\right) ^{2}}{2\sigma ^{2}_{t}}\right]
\end{equation}
where $A_t=\sigma_0\left(1+\frac{i\hbar t}{2m_0 \sigma^2_0}\right)$
and $\sigma_t=\left|A_{t}\right|=\sigma_0 \left(1+\frac{\hbar^2 t^{2}}{4m_0^2
\sigma^{4}_{0}}\right)^{1/2}$; $\sigma_t$ is the width of the wave packet 
at any instant $t$. \\

Next, we consider the spin part of the total wave function. 
As mentioned earlier, the initial spin of a spin-1/2 neutral particle is taken
to be polarized along the $+\widehat{\bf x}$-- axis; i.e., the initial
spin state is given by
\begin{equation}
\label{18}
\chi \left( 0\right) =\left| \rightarrow \right\rangle _{x}=\frac{1}{\sqrt{2}}\left[ \left| \uparrow \right\rangle _{z}+\left| \downarrow \right\rangle _{z}\right]
\end{equation}
and hence
\begin{equation}
\label{19}
{\bf s}\left( 0\right) =\frac{\hbar }{2}\chi ^{\dagger }\left( 0\right) {\bm \sigma }\chi \left( 0\right) =\frac{\hbar }{2}\widehat{\bf x}
\end{equation}
Then we proceed to calculate the time evolved spin part of the total
wave function under the interaction Hamiltonian $H=\mu {\bm \sigma }.{\bf B}$.
For this purpose we note that the constant magnetic field ${\bf B}=B\widehat{\bf z}$
within the SR, confined between $x=0$ and $x=d$, is directed
along the $+\widehat{\bf z}$-- axis. Then the time evolved spin state
$\chi \left( t\right)$ is given by
\begin{equation}
\label{20}
\chi \left( t\right) =\exp \left( \frac{-iHt}{\hbar }\right) \chi \left( 0\right) =\frac{1}{\sqrt{2}}e^{-i\omega t}\, \left[ \left| \uparrow \right\rangle _{z}+e^{2i\omega t}\left| \downarrow \right\rangle _{z}\right]
\end{equation}
whence
\begin{equation}
\label{21}
{\bf s}\left( t\right) =\frac{\hbar }{2}\chi ^{\dagger }\left( t\right) 
{\bm \sigma }\chi \left( t\right) =\frac{\hbar }{2}\left( \cos 2\omega t\, 
\widehat{\bf x}+\sin 2\omega t\, \widehat{\bf y}\right)
\end{equation}
where \( \omega =\mu B/\hbar  \). Given the above expressions for
\( \psi \left( x,t\right)  \), \( \rho \left( x,t\right)  \) and
\( {\bf s}\left( t\right)  \) corresponding to Eqs. (16),
(17) and (21) respectively, one can now calculate the total probability
current density at any given point from Eq.(14). We specifically evaluate
it at the exit point ${\bf X}(x=d,y=0,z=0)$ of the SR. Then the expression
for the total current density $ {\bf J}({\bf X},t)$ reduces to the form given by
\begin{eqnarray}
&&{\bf J}({\bf X},t)\\
\nonumber
&=&\rho(x=d,t) \left\{ u+\frac{\left( d-ut\right) \hbar ^{2}t}{4m_{0}^{2}
\sigma ^{4}_{0}+\hbar ^{2}t^{2}}\right\} \widehat{\bf x}\\
\nonumber
&+&\rho(x=d,t)\left\{ \frac{\hbar \left( ut-d\right) }{2m_{0}\sigma ^{2}_{t}}\sin 2\omega 
t\right\} \widehat{\bf z}
\end{eqnarray}
The first term in Eq.(22) is the usual Schroedinger current, and the
second term represents the additional contribution arising from the
spin of the particle. Next, substituting \( t=\phi /2\omega  \) in Eq.(22), 
one gets the following expression for the probability distribution of spin 
orientations for the particles emerging from the SR given by
\begin{eqnarray}
{\bf J}\left({\bf X}, \phi \right)
\nonumber
&=&\rho \left( d,\phi \right) \left\{ u+\frac{\left( d-\frac{u\phi }{2\omega }\right) 
\frac{\hbar ^{2}\phi }{2\omega }}{4m_{0}^{2}\sigma ^{4}_{0}+\frac{\hbar ^{2}\phi 
^{2}}{4\omega ^{2}}}\right\} \widehat{\bf x}\\
\nonumber
&+&\rho \left( d,\phi \right) \left\{ \frac{\hbar \left( \frac{u\phi }{2\omega }-d
\right) }{2m_{0}\sigma ^{2}_{\phi }}\sin \phi \right\} \widehat{\bf z}\\
&\equiv& {\bf J}_{Sch}\left( d,\phi \right) +{\bf J}_{Spin}
\left( d,\phi \right)
\end{eqnarray}
 where \( \rho \left( d,\phi \right) =\frac{1}{\left( 2\pi \sigma ^{2}_{\phi }\right) ^{1/2}}\exp \left\{ -\frac{\left( d-\frac{u\phi }{2\omega }\right) ^{2}}{2\sigma ^{2}_{\phi }}\right\}  \)
and $\sigma _{\phi }=\sigma _{0}\left[ 1+\frac{\hbar ^{2}\phi ^{2}}{16m^{2}_{0}
\omega ^{2}\sigma ^{4}_{0}}\right] ^{1/2}$.
Thus the probability density function corresponding to the distribution
of spins emerging from the SR is given by 
\begin{equation}
\label{24}
\Pi \left( \phi \right)=\frac{\left|{\bf J}\left({\bf X},\phi \right) \right|}
{\int_{0}^{2\pi }\left|{\bf J}\left({\bf X},\phi \right) \right| d\phi }
\end{equation}
where $\left|{\bf J}\left({\bf X},\phi \right) \right|$
is calculated from Eq.(23). Next, we proceed to present the results of a 
few numerical estimates for the observable probabilities 
$ P_{+}\left( \theta \right)$ and $ P_{-}\left( \theta \right)$ determined by 
Eqs. (4) and (5) respectively, where \( \Pi \left( \phi \right)  \) is calculated
using Eqs. (23) and (24).

\vskip 0.4cm 
Numerical estimates for $P_{+}(\theta)$ and $P_{-}(\theta)$.---These 
estimates have been done using different values of the relevant 
parameters; viz. the initial width \( (\sigma _{0}) \), the spatial 
extension \( (d) \) of the constant magnetic field confined within the SR, 
the group velocity \( (u) \) of the peak of the wave packet and the magnitude 
of the constant magnetic field \( (B) \).\\ 
                                                                                          
The choices of these parameters in our calculations are constrained
by the condition that the parameter \( d \) should be sufficiently
large compared to the half width of the wave packet peaked at the
exit point \( (x=d) \) of the SR; also, the relevant parameters \( d \)
, \( u \) and \( B \) are chosen such that the spin rotation angle
\( \phi  \) is effectively confined between \( \phi =0 \) and \( \phi =\pi  \)
so that the SG magnet does \emph{not} come in the way of the neutron
beam.\\

Now, for presenting the results of our estimates, we first show a
few representative curves (Figs. 2a and 2b) which correspond to the
\emph{probability density functions} \( \Pi \left( \phi \right)  \)
of the spin orientations of the particles emerging from the SR. Note
that the curves in Figs. 2a and 2b are respectively associated with
two different sets of choices (say, I and II) of the parameters \( d \),
\( u \), and \( B \). While the two curves in Fig. 2a correspond
to two different values of \( \sigma _{0} \) (\( 10^{-5}and \) \( 10^{-4} \)),
the two curves in Fig. 2b correspond to those values of \( \sigma _{0} \)
. The curves in Fig. 2a for the set I correspond to \( d=1cm \),
\textbf{\( u=3\times 10^{5}cm/s \)} and \textbf{\( B=10\, gauss \)},
while the curves in Fig. 2b for the set II correspond to \( d=2cm \),
\textbf{\( u=3\times 10^{5}cm/s \)} and \textbf{\( B=10\, gauss \)}.\\

Note that both the curves in Fig.2a represent the probability density
function \( \Pi \left( \phi \right)  \) peaked at \( \phi =\phi _{1}=
34.94767^{\circ } \), while the curves in Fig.2b represent \( \Pi \left( 
\phi \right)  \) peaked at \( \phi =\phi _{2}=69.8953^{\circ } \). For 
different choices of the initial width (\( \sigma _{0} \)) with the other 
relevant parameters \( d \) , \( u \) and \( B \) remaining fixed, it is
seen from each of Figs. 2a and 2b that the qualitative nature of these
curves and the location of their peak remain the \emph{same,} while
their variances \emph{differ, increasing} with \emph{decreasing} values
of \emph{\( \sigma _{0} \)}. 
\vskip 0.4cm

\begin{figure}[t]
{\rotatebox{0}{\resizebox{8.0cm}{6.0cm}{\includegraphics{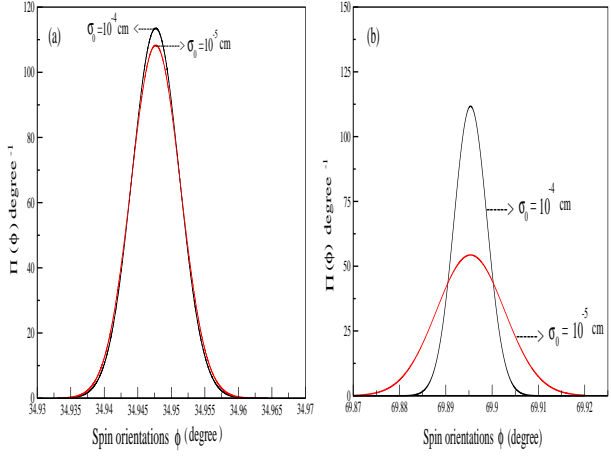}}}}
\caption{\label{fig.2} {\footnotesize The quantity $\Pi(\phi)$ denotes 
the probability density function which represents the distribution of spin 
orientations of the particles emerging from the SR. This quantity is plotted against
the angle of spin rotation $\phi$. The two curves in Fig.2a correspond 
to the set I of the choices \( d=1cm \), \( u=3\times 10^{5}cm/s \) and
\( B=10~ gauss \); both these curves are peaked at \( \phi =\phi _{1}=
34.94767^{\circ } \). The two curves in Fig.2b correspond to the set II 
of the choices \( d=2cm \), \( u=3\times 10^{5}cm/s \) and 
\( B=10\, gauss \); both these curves are peaked at \( \phi =\phi _{2}=
69.89534^{\circ } \).}}
\end{figure}
                                                       
Next, a few representative results of the numerical computations based
on using Eqs. (4), (5) and (24) are given in Tables 1 and 2 which
correspond respectively to the sets of values I and II of the parameters
\( d \) ,\( u \) and \( B \) (corresponding to the Fig. 2a and Fig. 2b 
respectively). The results shown in any one of these Tables indicate {\it how}
the observable quantities \emph{\( P_{+}\left( \theta \right)  \)}
and \emph{\( P_{-}\left( \theta \right)  \)} {\it vary} with different
initial widths \( \sigma_{0} \), corresponding to a specific orientation 
\( \left( \theta \right)  \) of the inhomogeneous magnetic field in the 
SG device (Fig.1). \\

It is seen that for a given value of $\theta$, for both the sets I and II of
the choices of the relevant parameters \( d \) ,\( u \) and \( B \), the
variation in the values of the probabilities $P_{+}(\theta)$ and $P_{-}(\theta)$
is very small, but is detectable for {\it smaller} values of $\sigma_0$. More
comprehensive estimates of $P_{+}(\theta)$, $P_{-}(\theta)$ and fluctuations
in the distribution of spins emerging from the SR, for a wider variation of
$\theta$ and other relevant parameters \( d \) ,\( u \) and \( B \), will be
presented in a later study.

\vskip 0.4cm
Table. 1. ~{\footnotesize The quantities} \emph{\footnotesize \( P_{+}\left(
\theta \right)  \)}{\footnotesize and} \emph{\footnotesize \( P_{-}\left( \theta
\right)  \)}{\footnotesize denote probabilities of the spin measurement along
different directions \( \widehat{n}(\theta ) \) making angles \( \theta  \)
with the initial spin polarised along \( +\widehat{x} \) - axis.
The numerical values of} \emph{\footnotesize \( P_{+}\left( \theta \right)  \)}
{\footnotesize and} \emph{\footnotesize \( P_{-}\left( \theta \right)  \)}
{\footnotesize are calculated for different initial widths \( \sigma _{0} \)
of the Gaussian wave packet. In this Table, the results are presented
for three different values of \( \theta  \) for the set I of the
values of the other relevant parameters \( d=1cm \),} \textbf{\footnotesize
\( u=3\times 10^{5}cm/s \)} {\footnotesize and} \textbf{\footnotesize
\( B=10\, gauss \),} {\footnotesize while} \textbf{\footnotesize \( \phi _{1}
=34.94767^{\circ } \)} {\footnotesize at which the curve of \( \Pi \left( \phi
\right)  \) is peaked}.\textbf{}\\

\begin{tabular}{|c|c|c|c|c|c|c|}
\hline
&\multicolumn{2}{c|}{$\theta=\phi_1$}&\multicolumn{2}{c|}{$\theta$}&\multicolumn{2}{c|}{$\theta$} \\
&\multicolumn{2}{c|}{$=34.94767^\circ$}&\multicolumn{2}{c|}{$=\phi_1+60^\circ$} & \multicolumn{2}{c|}{$=\phi_1+90^\circ$} \\
\hline
$\sigma_0$ & $P_{+}(\theta)$ & $P_{-}(\theta)$ & $P_{+}(\theta)$ & $P_{-}(\theta)$ &
$P_{+}(\theta)$ & $P_{-}(\theta)$\\
(cm)& & & & & & \\
\hline
$10^{-5}$&1.00000&0.00000&0.75000&0.25000&0.50000&0.50000\\
\hline
$10^{-6}$&1.00000&0.00000&0.75000&0.25000&0.50000&0.50000 \\
\hline
$10^{-7}$&0.99998&0.00002&0.75002&0.24998&0.50003&0.49997 \\
\hline
$10^{-8}$&0.99886&0.00114&0.75242&0.24758&0.50345&0.49655 \\
\hline
\end{tabular}
\vskip 0.4cm
                                                                                  
Table. 2.~{\footnotesize The quantities} \emph{\footnotesize \( P_{+}\left( \theta \right)  \)}{\footnotesize and} \emph{\footnotesize \( P_{-}\left( \theta \right)  \)}
{\footnotesize denote probabilities of the spin measurement along
different directions \( \widehat{n}(\theta ) \) making angles \( \theta  \)
with the initial spin polarised along \( +\widehat{x} \) - axis.
The numerical values of} \emph{\footnotesize \( P_{+}\left( \theta \right)  \)}
{\footnotesize and} \emph{\footnotesize \( P_{-}\left( \theta \right)  \)}
{\footnotesize are calculated for different initial widths \( \sigma _{0} \)
of the Gaussian wave packet. In this Table, the results are presented
for three different values of \( \theta  \) for the set II of the
values of the other relevant parameters \( d=2cm \),} \textbf{\footnotesize \( u=3\times 10^{5}cm/s \)}
{\footnotesize and} \textbf{\footnotesize \( B=10\, gauss \)}{\footnotesize ,
while \( \phi _{2}=69.89534^{\circ } \)at which the curve of \( \Pi \left( \phi \right)  \)
is peaked.}\\

\vskip 0.4cm 
\begin{tabular}{|c|c|c|c|c|c|c|}
\hline
&\multicolumn{2}{c|}{$\theta=\phi_2$}&\multicolumn{2}{c|}{$\theta$}&\multicolumn{2}{c|}{$\theta$} \\
&\multicolumn{2}{c|}{$=69.89534^\circ$}&\multicolumn{2}{c|}{$=\phi_2+60^\circ$} & \multicolumn{2}{c|}{$=\phi_2+90^\circ$} \\
\hline
$\sigma_0$ & $P_{+}(\theta)$ & $P_{-}(\theta)$ & $P_{+}(\theta)$ & $P_{-}(\theta)$ &
$P_{+}(\theta)$ & $P_{-}(\theta)$\\
(cm)& & & & & & \\
\hline
$10^{-5}$&1.00000&0.00000&0.75000&0.25000&0.50000&0.50000\\
\hline
$10^{-6}$&1.00000&0.00000&0.75000&0.25000&0.50000&0.50000 \\
\hline
$10^{-7}$&0.99995&0.00005&0.75004&0.24996&0.50006&0.49994 \\
\hline
$10^{-8}$&0.99546&0.00454&0.75355&0.24645&0.50672&0.49328 \\
\hline
\end{tabular}

\vskip 0.4cm

Summary and Outlook.---In the setup discussed in this paper, the 
\emph{observable quantities} are the probabilities \( P_{+}\left( \theta \right)  \)
and \( P_{-}\left( \theta \right)  \) which correspond to the measurement
of a spin variable along any direction by a SG device performed on
the particles emerging from the SR. Evaluations of these quantities
crucially depend on the probability distribution \( \Pi \left( \phi \right)  \)
of the orientations of spins of the particles emerging from the SR. This
in turn depends on the quantity \( \Pi \left( t\right)  \) which
corresponds to the distribution of transit times over which the particles
interact with the magnetic field while passing through the SR. \\

The quantity \( \Pi \left( t\right)  \) is calculated in this paper
for spin-1/2 particles in terms of the modulus of the probability
current density. Hence the estimates presented here for the observable
probabilities \( P_{+}\left( \theta \right)  \) and \( P_{-}\left( \theta \right)  \)
are ultimately determined by the modulus of the probability current
density. Thus if the experimental results for such a setup corroborate
such predictions, this would constitute a verification of the \emph{observability}
of the \emph{probability current density}. \\
                                                                                                         
As mentioned earlier, beacuse of an inherent \emph{nonuniqueness},
there are also \emph{other} quantum mechanical approaches which can
be used to evaluate the time distribution \( \Pi \left( t\right)  \),
apart from the specific scheme we've used in this paper based on the
modulus of the probability current density. In the context of our
setup, it should be instructive to derive the respective predictions
for the observable probabilities \( P_{+}\left( \theta \right)  \),
\( P_{-}\left( \theta \right)  \) from these different approaches 
with different criteria for fixing the {\it initial time} at which
the spin part of the wave function {\it starts} interacting with the 
localized magnetic field. One can then compare those predictions with the 
results of the actual experiment. Therefore it seems that an experimental 
study of the example discussed in this paper could be a worthwhile effort. \\

An interesting question of both theoretical and of experimental relevance
is as to what effect on the analysis does the assumption of a Gaussian
wave packet have. It should definitely be instructive to study the 
situation taking a non-Gaussian wave function for the spatial part which
is proposed to be done as a sequel to this work, apart from other sequels
based on different criteria for fixing the {\it initial time} of interaction
and the {\it duration} over which a {\it propagating wave packet} interacts 
with the localised potential.

\vskip 0.5cm

{\bf Acknowledgments}               

Comments by Anthony Leggett and Rick Leavens concerning 
our earlier paper on the arrival time distribution helped in motivating the present
work. Questions arising from interactions with Paul Davies and
John Corbett while pursuing collaboration on a different problem were
also useful for this work. Thanks are due to Will Marshall for helpful
discussions while this paper was being written. We are also grateful
to Peter Holland, Rick Leavens and Juan Gonzalo Muga for the
critical questions they raised after reading the initial draft of
this paper. We thank the referees for their helpful suggestions that led
to the present version of the paper.\\
                                                                                                         
DH acknowledges the support of the Jawaharlal Nehru Fellowship. AKP
and MMA acknowledge respectively the Junior Research Fellowship and
Senior Research Fellowship of the CSIR, India.

\end{document}